\documentclass[prl,twocolumn]{revtex4}
\usepackage{wasysym}
\usepackage{epsf}
\usepackage{graphicx}
\usepackage{bm}
\begin{document}
\title{Experimental evidence for magnetorotational
instability in a helical magnetic field}

\author{Frank Stefani, Thomas Gundrum, Gunter Gerbeth}
\affiliation{Forschungszentrum Rossendorf, 
P.O. Box 510119, D-01314 Dresden, Germany}
\author{G\"unther R\"udiger, Manfred Schultz, Jacek Szklarski}
\affiliation{Astrophysikalisches Institut Potsdam, 
An der Sternwarte 16, D-14482 Potsdam, Germany}
\author{Rainer Hollerbach}
\affiliation{Department of Applied Mathematics, University
of Leeds, Leeds, LS2 9JT, United Kingdom}


\begin{abstract}
A recent paper [R. Hollerbach and  G. R\"udiger, Phys. Rev. Lett. 
{{\bf 95}}, 124501 (2005)] has shown that the threshold for the onset
of the magnetorotational instability (MRI) in a Taylor-Couette flow 
is dramatically reduced if both axial and azimuthal magnetic fields
are imposed. In agreement with this prediction, we present results of
a Taylor-Couette experiment with the liquid metal alloy GaInSn,
showing evidence for the existence of the MRI at Reynolds numbers of
order 1000 and Hartmann numbers of order 10.
\end{abstract}


\maketitle

The role of magnetic fields in the cosmos is two-fold:  First, planetary,
stellar and galactic fields are a product of the homogeneous dynamo effect
in electrically conducting fluids. Second, magnetic fields are also
believed to play an active role in cosmic structure formation, by enabling
outward transport of angular momentum in accretion disks via the
magnetorotational instability (MRI) \cite{BAHA}.  Considerable theoretical
and computational progress has been made in understanding both processes.
The dynamo effect has even been verified experimentally, in large-scale
liquid sodium facilities in Riga and Karlsruhe, and continues to be
studied in laboratories around the world \cite{RMP}.  In contrast,
obtaining the MRI experimentally has been less successful thus far
\cite{ROSNER}. (\cite{LATHROP} claim to have observed it, but their
background state was already fully turbulent, thereby defeating the
original idea that the MRI would destabilize an otherwise stable flow.)

\begin{figure}[ht]
\begin{center}
\epsfxsize=8.6cm\epsfbox{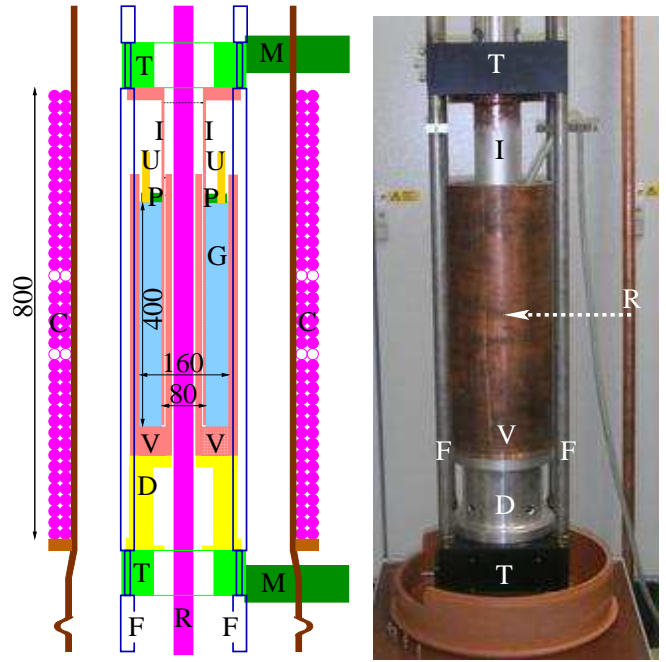}
\vspace*{4mm}
\caption{Sketch (left) and photograph (right) of the central 
module of the PROMISE facility.
V - Copper vessel, I - Inner cylinder, G - GaInSn, 
U - Two ultrasonic transducers, 
P - Plexiglass lid,
T - High precision turntables, 
M - Motors,
F - Frame, 
C - Coil,
R - Central rod. The dimensions are in mm.  When the experiment is running,
the rod R in the right picture goes through the center of the cylinders.}
\end{center}
\end{figure}

If only an axial magnetic field is externally applied, the azimuthal
field that is necessary for the occurrence of the MRI must be produced
by induction effects, which are proportional to the magnetic 
Reynolds number ($Rm$) of the flow.  But why not substitute this
induction process simply by {\it externally applying an azimuthal
magnetic field as well\/}? This question was at the heart of the paper
\cite{HORU}, where it was shown that the MRI is then possible with
far smaller Reynolds ($Re$) and  Hartmann ($Ha$) numbers.  In this
paper we report experimental verification of this idea, presenting
evidence of the MRI in a liquid metal Taylor-Couette (TC) flow with
externally imposed axial and azimuthal (i.e., helical) magnetic fields.

The heart of our facility `PROMISE' ({\it P\/}otsdam {\it RO\/}ssendorf
{\it M\/}agnetic {\it I\/}n{\it S\/}tability {\it E\/}xperiment) is a
cylindrical vessel made of copper (see Fig. 1). The use of copper was
motivated by the fact that the instability usually occurs at lower Reynolds
and Hartmann numbers for the case of ideally conducting boundaries than for
non-conducting boundaries \cite{AN}.  The inner wall is 10 mm thick, and
extends in radius from 22 to 32 mm; the outer wall is 15 mm thick,
extending from 80 to 95 mm in $r$.  This vessel is filled with the alloy
Ga$^{67}$In$^{20.5}$Sn$^{12.5}$, which is liquid at room temperatures.
The physical properties of GaInSn at 25 $^{\circ}$C are: density
$\rho=6.36 \times 10^3$ kg/m$^3$, kinematic viscosity $\nu=3.40\times
10^{-7}$ m$^2$/s, electrical conductivity $\sigma=3.27\times 10^6$
($\Omega$ m)$^{-1}$. The magnetic Prandtl number is then $Pm=\mu_0
\sigma \nu=1.40 \times 10^{-6}$.

\begin{figure}[!ht]
\begin{center}
\epsfxsize=6.2cm\epsfbox{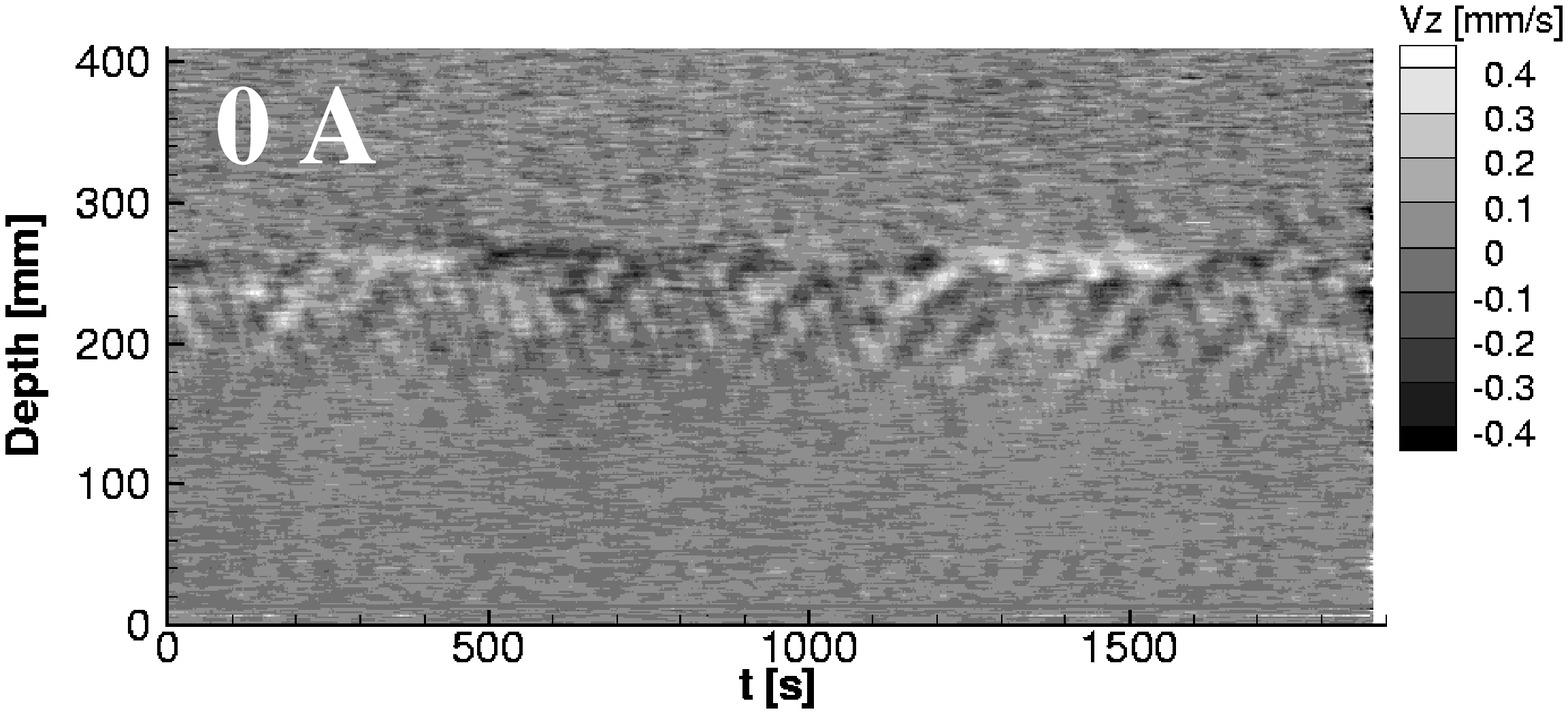}
\epsfxsize=6.2cm\epsfbox{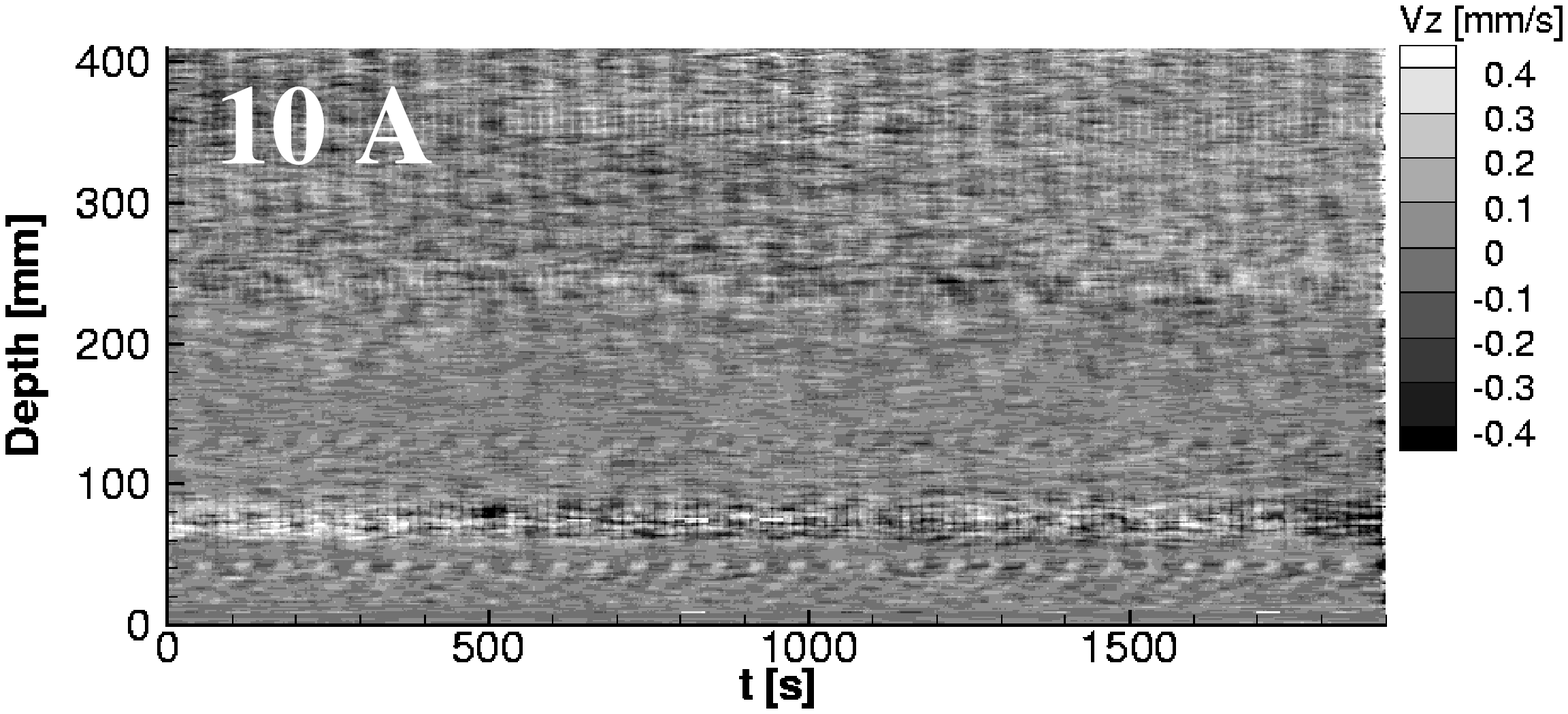}
\epsfxsize=6.2cm\epsfbox{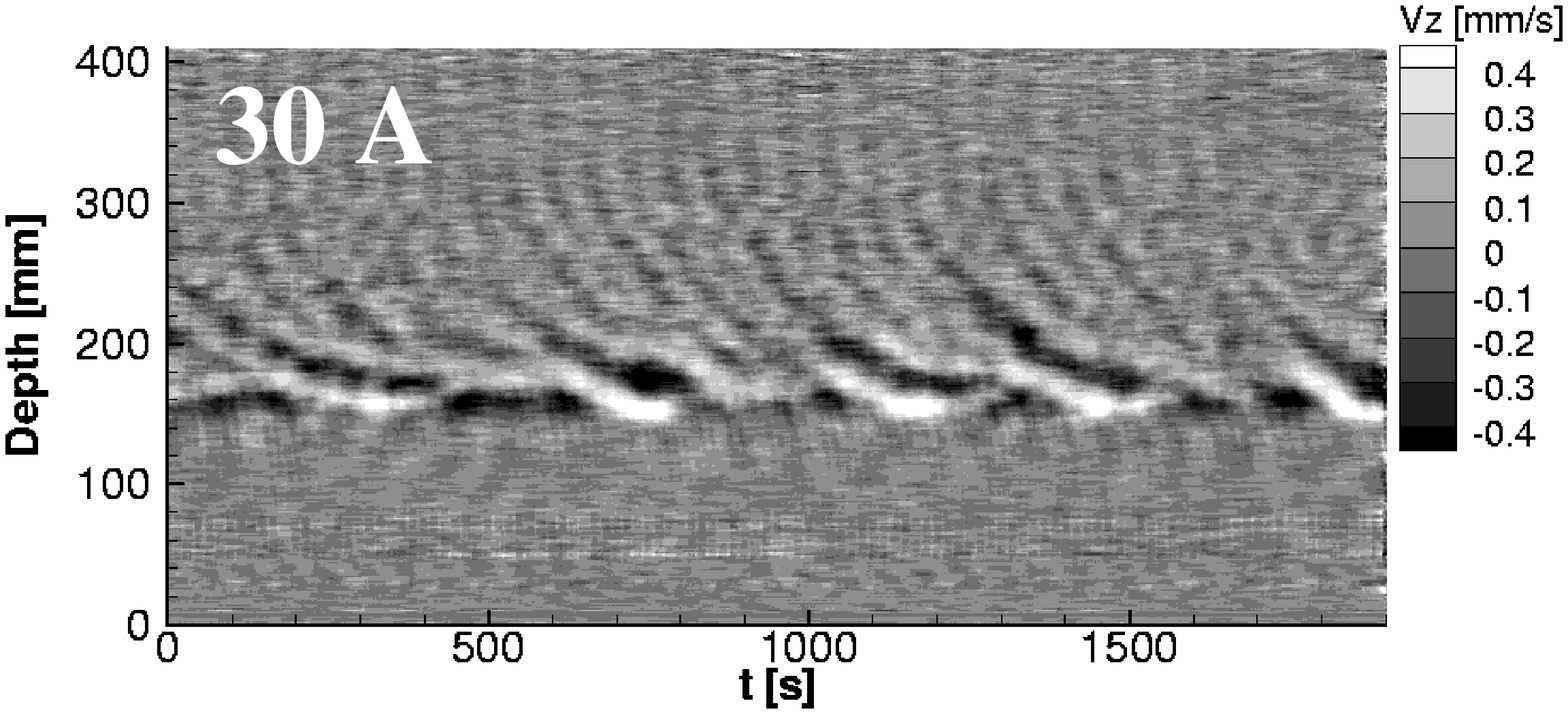}
\epsfxsize=6.2cm\epsfbox{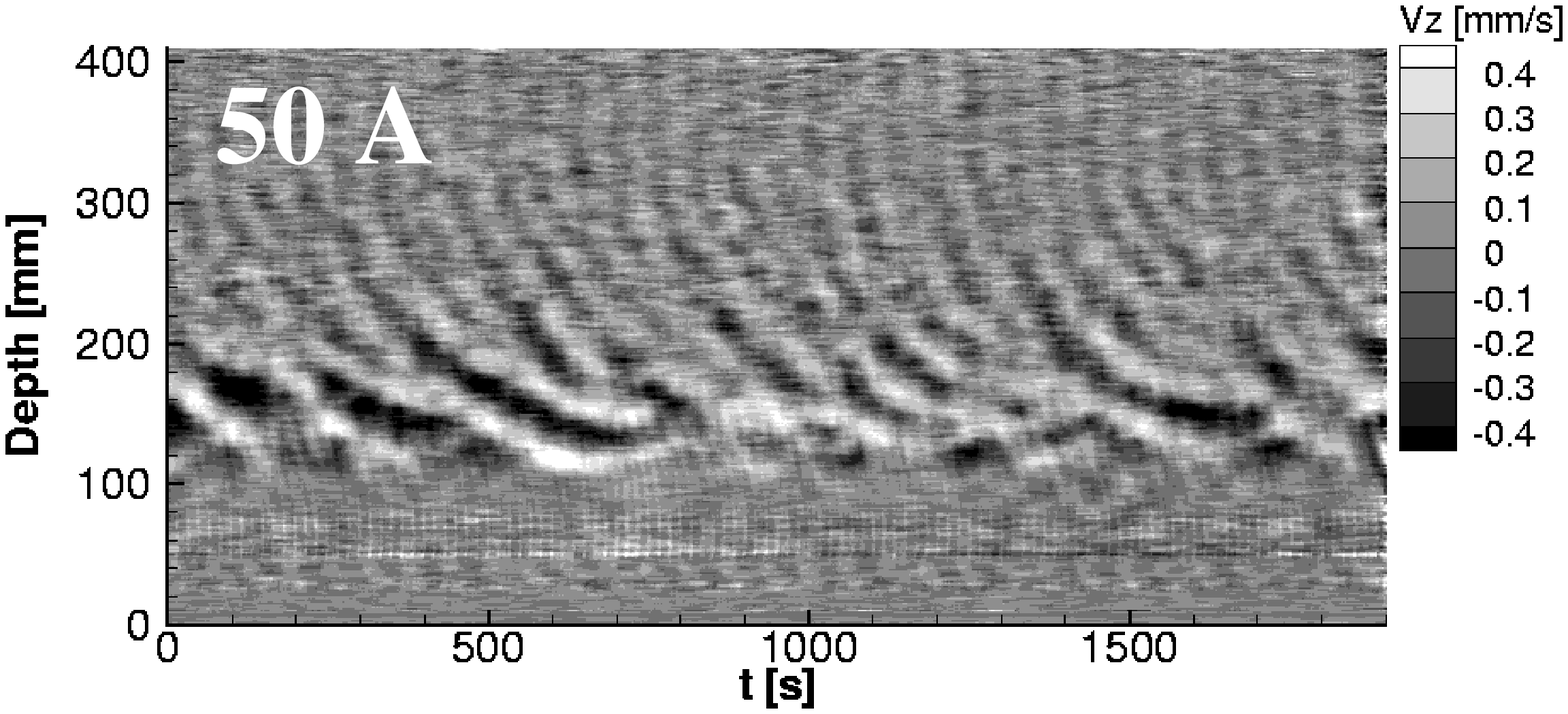}
\epsfxsize=6.2cm\epsfbox{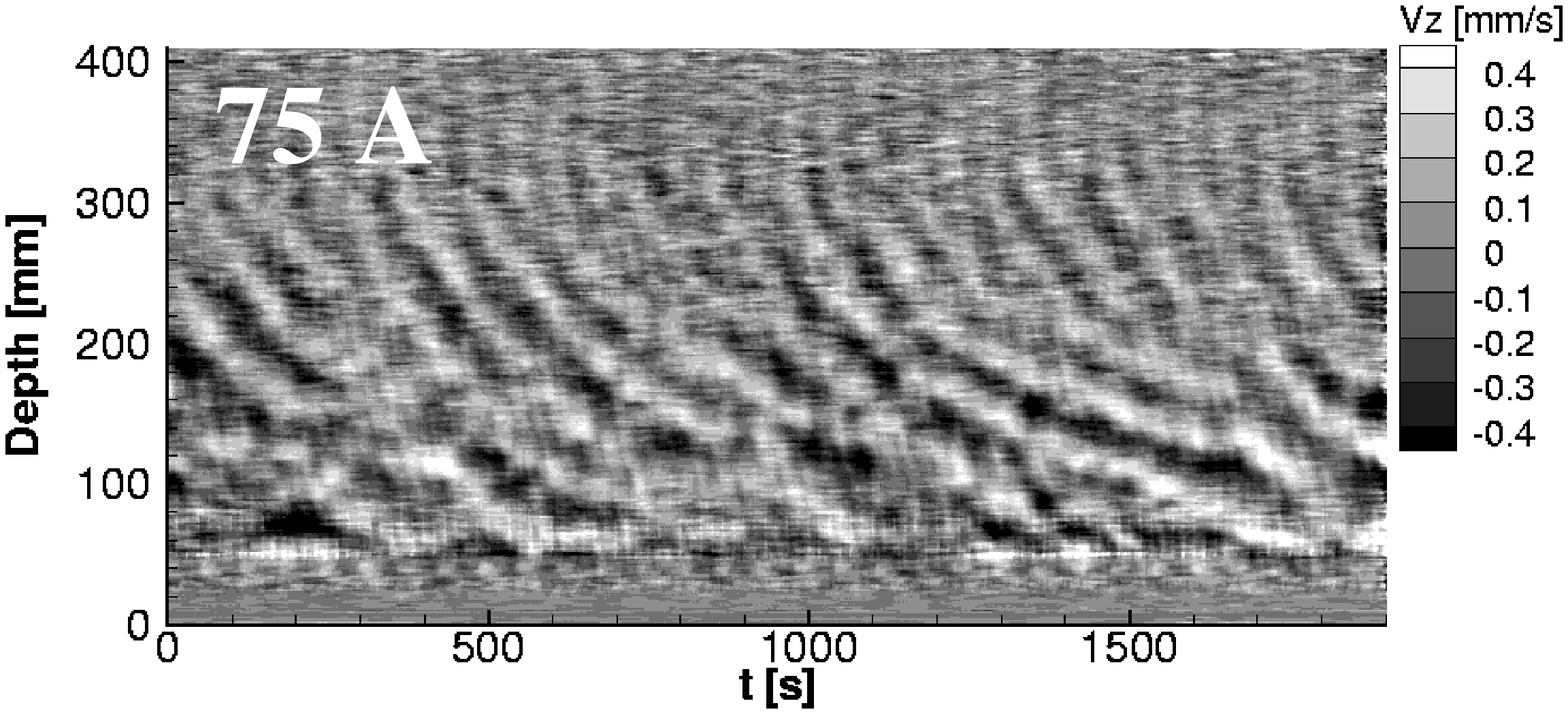}
\epsfxsize=6.2cm\epsfbox{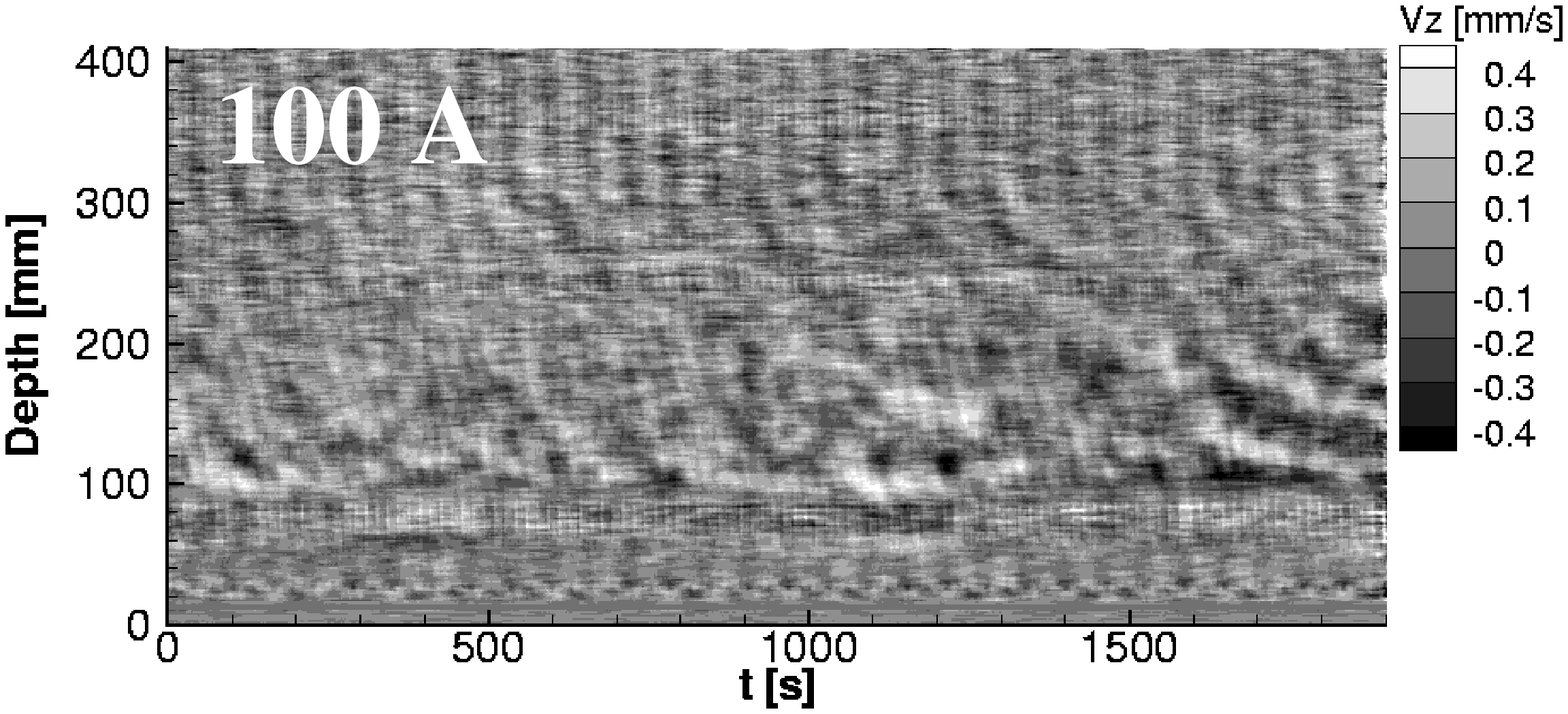}
\epsfxsize=6.2cm\epsfbox{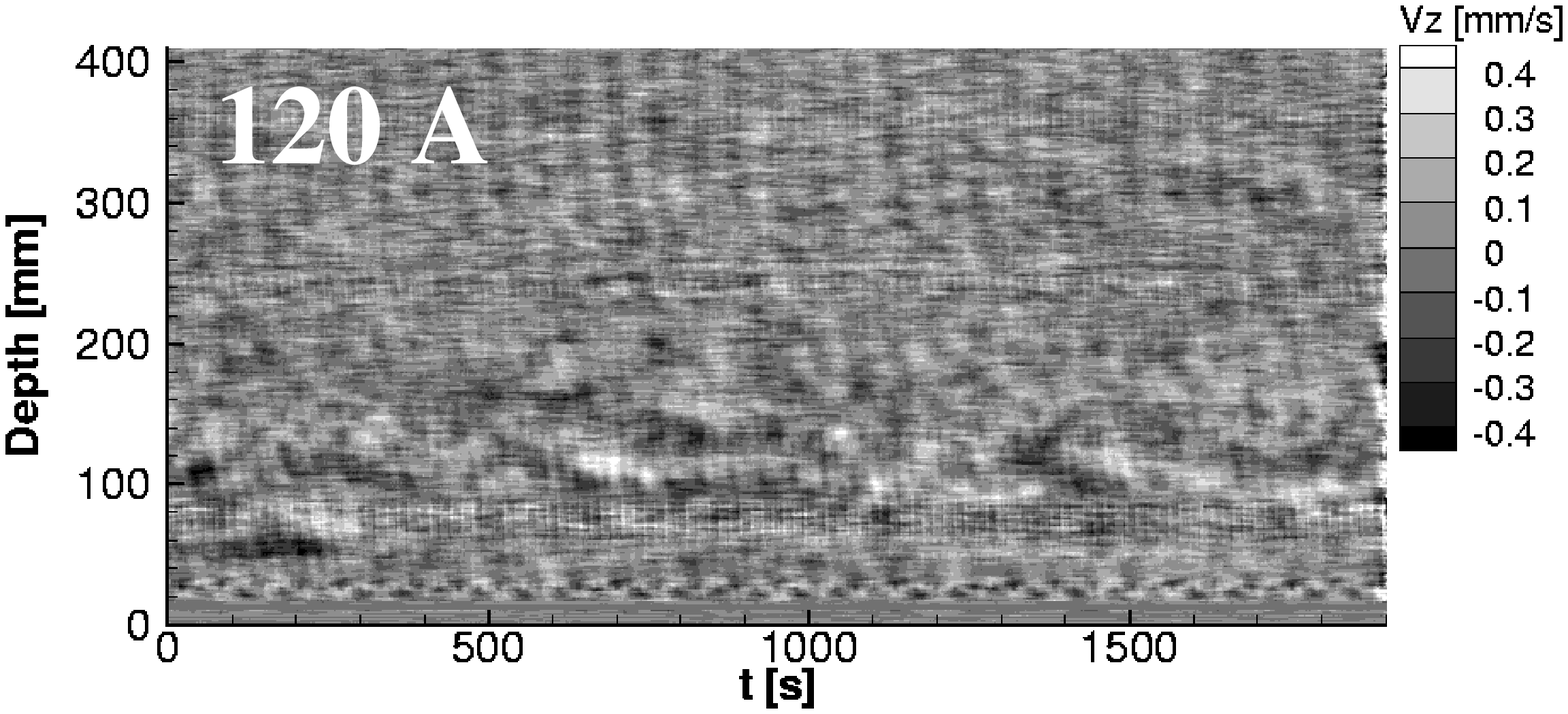}
\caption{Measured axial velocity in dependence on time and depth  
for different coil currents.}
\end{center}
\end{figure}

The copper vessel is fixed, via an aluminum spacer D, on a precision 
turntable T; the outer wall of the vessel thus serves as the outer
cylinder of the TC flow. The inner cylinder of the TC flow is fixed to
an upper turntable, and is immersed into the GaInSn from above.  It has
a thickness of 4 mm, extending in radius from 36 to 40 mm.  There is thus
a gap of 4 mm between this immersed cylinder and the inner wall of the
containment vessel.  The actual TC flow then extends between 
$r_{\rm in}=40$ mm and $r_{\rm out}=80$ mm.  At the moment, the upper 
endplate is a
plexiglass lid P fixed to the frame F. In contrast, the bottom is simply
part of the copper vessel, and hence rotates with the outer cylinder.
There is thus a clear asymmetry in the endplates, with respect to both
their rotation rates and electrical conductivities.

Note also that the accuracy of this setup is not quite at the $\sim\!10^{-2}$
mm level that can be achieved in ordinary, hydrodynamic TC experiments,
e.g.  \cite{SCHULTZGRUNOW}.  For example, in order to ensure a well-defined
electrical contact between the GaInSn and the walls, it is necessary
to intensively rub the fluid into the copper.  The precision is then more
like $10^{-1}$ mm.  The Reynolds number $Re=2 \pi f_{\rm in} r_{\rm in}
(r_{\rm out}-r_{\rm in})/\nu$ is also $O(10^3)$, considerably greater than
the $Re_c=68$ value obtained in the nonmagnetic problem (with stationary
outer cylinder).

Axial magnetic fields of order 10 mT are produced by a double-layer coil 
(C). The omission of windings at two symmetric positions close to the middle
resulted from a coil optimization to maximize the homogeneity of the axial
field  throughout the volume occupied by the liquid. This coil is fed by a
power supply that can deliver up to 200 A. The azimuthal field, also
of order 10 mT, is  generated by a current through a water-cooled copper
rod R of radius 15 mm. The power supply for this axial current delivers up
to 8000 A.

\begin{figure}[ht]
\begin{center}
\epsfxsize=8.6cm\epsfbox{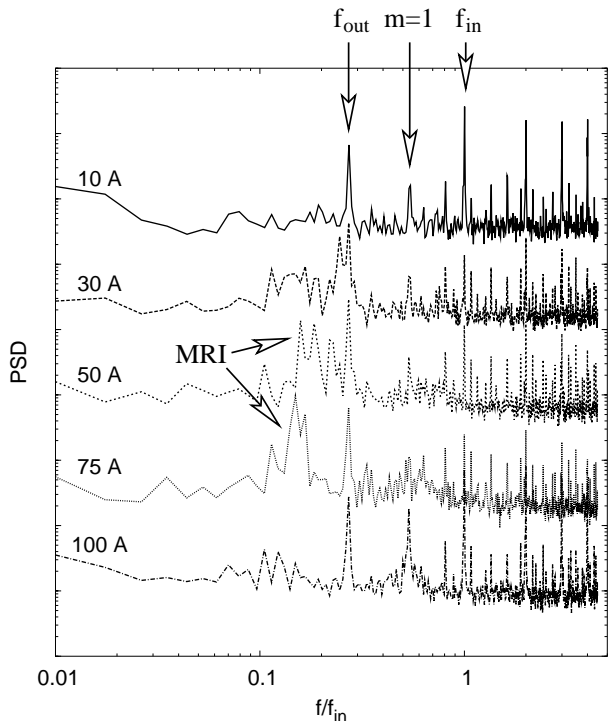}
\vspace{2mm}
\caption{Power spectra of the $v_z$  fluctuations, averaged over the depths
between 220 and 308 mm, for 5 different coil currents. The frequencies of the
inner and  outer cylinder are indicated. The MRI is visible at 50 and 75
A. In addition to that, an $m=1$ mode is also present.}
\end{center}
\end{figure}

At present, the measuring instrumentation consists exclusively of two
ultrasonic transducers with a working frequency of 4 MHz; these are fixed
into the plexiglass lid, 15 mm away from the outer copper walls, flash
mounted at the interface to the GaInSn.  They provide full axial velocity
profiles along the ultrasound beamline by means of Ultrasound Doppler
Velocimetry (UDV) \cite{CRAMER}. A similar measuring system for the axial
velocity components in TC experiments was first described by Takeda
\cite{TAKEDA}. The comparison of the two signals from two opposite sensors 
is important in order to clearly distinguish between the expected 
axisymmetric ($m=0$) MRI mode \cite{AN} and certain $m=1$ modes which also
play a role in some parameter regions of the experiment.

Typically, the duration of the experimental runs was 1900 sec, after a
waiting time of one hour. Such a long waiting time was chosen not only due
to the hydrodynamic gap diffusion time, but also due to our numerical
predictions of rather small growth rates of the MRI mode in helical fields
(see also \cite{JI}).

One of the features of TC flow under the influence of helical magnetic
fields is the replacement of the Taylor vortex flow by a travelling wave
\cite{HORU,AN}.  This travelling wave appears already at $\mu:=f_{\rm out}
/f_{\rm in}=0$, although with a very low frequency. With increasing $\mu$,
the frequency increases and typically reaches a value of $0.2 f_{\rm in}$ at
the Rayleigh line $\mu_{\rm Ray}=(r_{\rm in}/r_{\rm out})^2=:\eta^2$
(we have $\eta=0.5$ here, hence $\mu_{\rm Ray}=0.25$).  The crucial point
now is that under the influence of helical magnetic fields, the critical
Reynolds numbers remain relatively small increasingly far beyond the
Rayleigh line, where nothing special happens.  Typically, this shift to
larger $\mu$ becomes larger for increasing values of the ratio $\beta:=
B_{\varphi}(r=r_{\rm in})/B_{z}$ of azimuthal field to axial field. 

Another typical feature of the MRI is that, for fixed $Re$, it sets in at
a certain critical value of the Hartmann number $Ha=B_{z} (r_{\rm in}
(r_{\rm out}-r_{\rm in})\sigma/\rho \nu)^{1/2}$, and disappears
again at some higher value. In this letter we focus exclusively on
experimental results which substantiate this behavior.

All results presented in the following are for rotation rates of $f_{\rm in}
=0.06$ s$^{-1}$ and $f_{\rm out}=0.0162$ s$^{-1}$, i.e. for $\mu=0.27$,
beyond the Rayleigh line $\mu_{\rm Ray}=0.25$.  Figure 2 documents a
selection of seven experimental runs for coil currents $I_{\rm coil}$ between
0 and 120 A. In each case, the axial current $I_{\rm rod}$ was fixed to 6000
A. The color coding of the plots indicates the (negative) axial velocity
component measured along the ultrasound beam, from which we have subtracted
the depth-dependent time average in order to filter out the two Ekman
vortices which appear already without any magnetic field.  These vortices,
characterized by inward radial flows close to the upper and lower endplates,
show up in our UDV data in the form of positive and negative axial velocities, 
separated approximately at mid-height by a rather sharp boundary.  This
boundary most likely corresponds to a jet-like radial outflow in the center
of the cylinder, as discussed in \cite{KAGE}.

For $I_{\rm coil}=30$ A we already observe a traveling wave-like 
structure which is, however, still restricted to the middle part of 
the cylinder. One might speculate that, due to the (jet-like) radial outward
flow there, fluid with lower angular momentum is transported outward, which
leads to a steeper decrease of $v_{\varphi}$ and hence to a (locally) lower 
value $\mu$ than what would be expected  from the rotation ratio of outer
to inner cylinder \cite{KAGE}.

For increasing $I_{\rm coil}$, this traveling wave becomes more and more
dominant, until at 75 A it fills essentially the entire cylinder. Increasing
$I_{\rm coil}$ even further though, at 100 A this wave ceases to exist, and
we again have a rather featureless flow.  This is shown more quantitatively
in Fig. 3, where we show the average over the depths 220-308 mm of the power
spectral density for the axial velocity for 5 selected values of
$I_{\rm coil}$. A feature common to all five curves is the appearance of
the rotation rates of the inner and outer cylinders. This reflects
certain geometric imperfections of the facility, probably in the form of
cylinder eccentricities or metal oxides sticking to some parts of the walls.
There is also a certain peak approximately at the mean of the inner and
outer cylinder frequencies, which, on closer inspection of the two
transducer signals, turns out to be a non-axisymmetric $m=1$ mode.
Most interesting for us here, however, is the MRI mode at $f/f_{\rm in}
\sim 0.1-0.2$. These frequencies have been analyzed in detail for all
values of $I_{\rm coil}$ chosen in the experiment.

\begin{figure}[ht]
\begin{center}
\epsfxsize=8.6cm\epsfbox{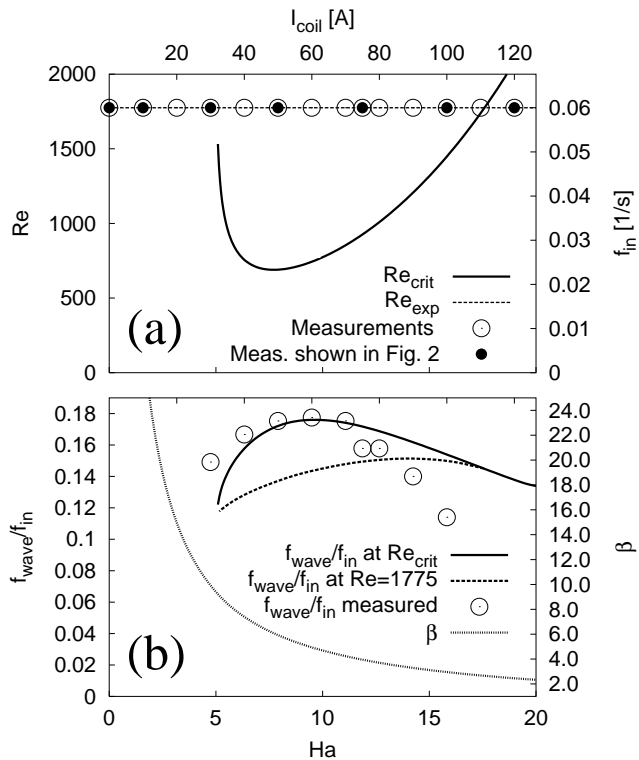}
\vspace{2mm}
\caption{(a) Computed critical Reynolds numbers for varying $I_{\rm coil}$
(and hence Hartmann numbers), for fixed $I_{\rm rod}=6000$ A.
(b) Frequency $f_{\rm wave}$ of the travelling MRI wave, normalized to the
rotation rate $f_{\rm in}$ of the inner cylinder. The ratio of azimuthal
field to axial field $\beta$ is also shown. }
\end{center}
\end{figure}

We now interpret these experimental data in the context of numerical
predictions, obtained and cross-checked by various independent codes
\cite{HORU,SHALYBKOV,AIP} for the solution of the linear eigenvalue problem
in unbounded cylinders. First, Fig.\ 4a shows the dependence of the critical
Reynolds number on the Hartmann number. We observe the typical MRI behavior,
with the minimum value of $Re_{\rm crit}$ occurring at $Ha=7.7$. Fig.\ 4b
compares the measured frequencies of the travelling wave, normalized to the
rotation rate of the inner cylinder, with the computed ones. The latter are
actually given in two versions, at $Re_{\rm crit}$, and at the experimental
value $Re=1775$.  It is interesting that the difference between the two
curves is relatively small. Hence it might be anticipated that the
normalized frequency in the nonlinear  (saturated) regime is probably also
not far from these values. The measured frequencies are in reasonable
correspondence with the computed ones, and show a similar behavior, with a
maximum close to $Ha=10$.

In summary, we have provided experimental evidence for the existence of the
MRI in current-free helical magnetic fields, by showing its appearance in a
certain interval of Hartmann numbers, in reasonable agreement with numerical
predictions. Further experimental results for other combinations of the
governing parameters will be published elsewhere.  Certainly, much work
remains to be done, including a detailed investigation into the role of the
magnetic axial  boundary conditions.  For later experiments, a
symmetrization of these boundaries is envisioned. Connected with this, a
suppression of the Ekman vortices by means of split rings (proposed in
\cite{KAGE}, see also \cite{HF}) may also help to avoid special effects in
the mid-height of the cylinder. 

This work was supported by the German Leibniz Gemeinschaft, within its SAW
program.  We thank R. Rosner for his interest and support for this work,
Heiko Kunath for technical assistance, and Markus Meyer for assistance
taking some of the data.  The Rossendorf group also thanks Janis Priede and
Ilmars Grants for stimulating discussions.

\end{document}